\documentclass[prl,twocolumn,showkeys,showpacs,preprintnumbers,amsmath,amssymb]{revtex4}

\usepackage{graphicx}
\usepackage{dcolumn}
\usepackage{bm}
\usepackage{times}

\def\tr{\textrm{tr}}

\def\cA{\mathcal{A}}
\def\cM{\mathcal{M}}

\def\ZZ{\mathbb{Z}}

\def\stamp{--- {\bf \today} --- {\bf \jobname.tex}}

\def\cG{\mathcal{G}}
\def\cB{\mathcal{B}}

\def\tr{\textrm{tr}}
\def\cSS#1#2{{\mathcal{S}}_{{#1},{#2}}}

\def\BE{\begin{equation}}
\def\EE{\end{equation}}
\def\spa#1.#2{\left\langle#1\,#2\right\rangle}
\def\spb#1.#2{\left[#1\,#2\right]}
\def\spba#1.#2.#3{\left[#1|#2|#3\right\rangle}
\def\spab#1.#2.#3{\left\langle#1|#2|#3\right]}
\def\spaa#1.#2.#3{\left\langle#1|#2|#3\right\rangle}
\def\spbb#1.#2.#3{\left[#1|#2|#3\right]}
\def\lor#1.#2{\left(#1\,#2\right)}

\def\Year{\expandafter\eatPrefix\the\year}
\newcount\hours \newcount\minutes
\def\monthname{\ifcase\month\or
January\or February\or March\or April\or May\or June\or July\or
August\or September\or October\or November\or December\fi}
\def\shortmonthname{\ifcase\month\orx
Jan\or Feb\or Mar\or Apr\or May\or Jun\or Jul\or
Aug\or Sep\or Oct\or Nov\or Dec\fi}

\def\TimeStamp{\hours\the\time\divide\hours by60%
\minutes -\the\time\divide\minutes by60\multiply\minutes by60%
\advance\minutes by\the\time%
${\rm \shortmonthname}\cdot   \if\day<10{}0\fi\the\day\cdot   \the\year
\qquad\the\hours:\if\minutes<10{}0\fi\the\minutes$}

\newskip\humongous \humongous=0pt plus 1000pt minus 100pt

\newif\ifdtup

\newcounter{eqnumber}[section]

\def\tr{\mathop{\rm tr}\nolimits}

\newbox\charbox
\newbox\slabox

\def\spa#1.#2{\left\langle#1\,#2\right\rangle}
\def\spb#1.#2{\left[#1\,#2\right]}
\def\lor#1.#2{\left(#1\,#2\right)}

\catcode`@=11  

\def\lsl{\not{\hbox{\kern-2.3pt $\ell$}}}
\def\ksl{\not{\hbox{\kern-2.3pt $k$}}}

\def\spa#1.#2{\left\langle#1\,#2\right\rangle}
\def\spb#1.#2{\left[#1\,#2\right]}
\def\lor#1.#2{\left(#1\,#2\right)}

\def\sand#1.#2.#3{%
  \left\langle\smash{#1}{\vphantom1}\right|{#2}%
  \left|\smash{#3}{\vphantom1}\right\rangle}
\def\sandp#1.#2.#3{%
  \left\langle\smash{#1}{\vphantom1}^{-}\right|{#2}%
  \left|\smash{#3}{\vphantom1}^{+}\right\rangle}
\def\sandpp#1.#2.#3{%
  \left\langle\smash{#1}{\vphantom1}^{+}\right|{#2}%
  \left|\smash{#3}{\vphantom1}^{+}\right\rangle}
\def\sandmm#1.#2.#3{%
  \left\langle\smash{#1}{\vphantom1}^{-}\right|{#2}%
  \left|\smash{#3}{\vphantom1}^{-}\right\rangle}
\def\sandpm#1.#2.#3{%
  \left\langle\smash{#1}{\vphantom1}^{+}\right|{#2}%
  \left|\smash{#3}{\vphantom1}^{-}\right\rangle}
\def\sandmp#1.#2.#3{%
  \left\langle\smash{#1}{\vphantom1}^{-}\right|{#2}%
  \left|\smash{#3}{\vphantom1}^{+}\right\rangle}

\def\tr{\mathop{\hbox{\rm tr}}\nolimits}

\begin{document}

\preprint{IHES-P-09-33,\ IPHT-T-09/092}

\title{Minimal Basis for Gauge Theory Amplitudes}

\author{N. E. J. Bjerrum-Bohr}
\author{Poul H. Damgaard}
\affiliation{Niels Bohr International Academy,\\ The Niels Bohr
Institute,
Blegdamsvej 17,\\ DK-2100, Copenhagen \O, Denmark}
\email{bjbohr@nbi,dk, phdamg@nbi.dk}

\author{Pierre Vanhove}
\affiliation{Institut des Hautes Etudes Scientifiques, Le Bois-Marie,\\
        F-91440 Bures-sur-Yvette, France\\  and\\
        CEA, DSM, Institut de Physique Th\'eorique, IPhT, CNRS, MPPU,\\
        URA2306, Saclay, F-91191 Gif-sur-Yvette, France}%
 \email{pierre.vanhove@cea.fr}

\date{\today}

\begin{abstract}
Identities based on monodromy for integrations in string theory
are used to derive relations between different color ordered
tree-level amplitudes in both bosonic and supersymmetric string theory.
These relations imply that the color ordered tree-level $n$-point gauge theory
amplitudes can be expanded in a minimal basis of $(n-3)!$ amplitudes.
This result holds for any choice of polarizations of the external
states and in any number of dimensions.
\end{abstract}

\maketitle

\section{Introduction}

The search for a consistent theoretical framework of particle
physics has led to remarkable progress in the understanding of
fundamental interactions in Nature. String theory provides a
very general unified language that naturally incorporates field
theories of phenomenological interest and gravity in the
low-energy limit. Much can be learned from studying the
organizational and computational inspiration it
poses~\cite{Green:1987sp}. One striking aspect is the link
string theory can provide between gravity and gauge theories.
Concrete examples of such relationships include the
Kawai-Lewellen-Tye~\cite{KLT} relations which connect
amplitudes in closed and open string theories. In the
low-energy limit this gives a very puzzling and non-trivial map
between perturbative amplitudes in gravity and Yang-Mills
theory that is far from obvious when viewed at the field theory
perspective~\cite{Bern:2002kj}.

In this Letter, we will consider a set of relations among
tree-level string theory amplitudes that are implied by their
defining integrals. Different color orderings of external legs
are connected to specific integration regimes on the string
world sheet, but they can be related to each other through
monodromy relations. In the field theory limit the phase
relations between different integrals induced by these
monodromy considerations reduce to a set of equations linking
gauge theory amplitudes with different color traces.
 We first remark that by cyclicity of
the trace the number of color ordered amplitudes is reduced from $n!$ to
$(n-1)!$ The full
set of monodromy relations for the color-ordered amplitudes
imply a drastic reduction of the number of independent
amplitudes in the $n$-point case. The number of basis
amplitudes is in this way reduced from $(n-1)!$ to $(n-3)!$
Analogously to the Kawai-Lewellen-Tye relations, the detailed
understanding of the underlying identities at the gauge theory
level poses an interesting challenge. The existence of a
minimal number of $(n-3)!$ basis amplitudes in gauge theory,
and an associated set of identities, has been conjectured by
Bern et al.~\cite{Bern:2008qj} (see also
ref.~\cite{Sondergaard:2009za} for the extension to gauge
theory with matter) and already checked explicitly to a high
number of external legs with different combinations of external
states and helicities. The origin of this reduction in basis amplitudes
appears in a particularly transparent manner from string
theory.

We will here briefly recall how to derive these
monodromy-induced relations for string theory amplitudes. The
$n$-point amplitude in open string theory with $U(N)$ gauge
group reads\vspace{-0.15cm}
\begin{eqnarray}
\label{e:AmpNdef}
\cA_n &=&ig_{\rm YM}^{n-2}\, (2\pi)^{D}\,\delta^{D}(k_1+\cdots+k_n)\,\\
&&\hspace{-1cm}\sum_{(a_1,\dots,a_n)\in S_n/\ZZ_n}\,\nonumber
\hspace{-1cm} \tr(T^{a_1}\cdots T^{a_n})\, \cA{(a_1,\cdots,a_n)}
\,,
\end{eqnarray}\vskip-6pt\noindent
where  $D$ is any number of dimensions obtained by dimensional
reduction from $26$ dimensions if we consider the bosonic
string, or $10$ dimensions in the supersymmetric case. In fact
our considerations are completely general and without reference
to any specific string theory. The color-ordered amplitudes on
the disc are given by~\cite{Green:1987sp}\vspace{-0.2cm}
\begin{eqnarray}
\label{eq:orderedBos}
\cA{(a_1,\cdots, a_n)}&\!\!=\!\!&\!\!\int\!\!\prod_{i=1}^{n}
dz_i\,{|z_{ab}\,z_{ac}\,z_{bc}|\over dz_adz_bdz_c}  \prod_{i=1}^{n-1}
H(x_{a_{i+1}}\!-\!x_{a_i}) \cr
&\times&\prod_{1\leq                i<j\leq               n}               \,
|x_i-x_j|^{2\alpha'k_i\cdot k_j}\, F_n\,,
\end{eqnarray}\vskip-6pt\noindent
with $dz_i=dx_i$ and  $z_{ij}=x_i-x_j$ for the bosonic case and
$dz_i=dx_i d\theta_i$ and $z_{ij}=x_i-x_j+\theta_i\theta_j$ for
the supersymmetric case. The ordering of the external legs is
enforced by the product of Heaviside functions such that
$H(x)=0$ for $x<0$ and $H(x)=1$ for $x\geq0$. The M\"obius
$SL(2,\mathbb{R})$ invariance requires one to fix the position
of three points denoted $z_a$, $z_b$ and $z_c$. A traditional
choice is $x_1=0$, $x_{n-1}=1$ and $x_n=+\infty$, supplemented
by the condition $\theta_{n-1}=\theta_n=0$ in the superstring
case.

All helicity dependence of the external states is contained in
the $F_n$ factor. For tachyons, one has $F_n=1$ while for $n$
gauge bosons with polarization vectors $h_i$ one arrives at
$F_n=\exp-\sum_{i\neq j}\big( {\sqrt{\alpha'} (h_i\cdot
k_j)\over (x_i-x_j)}-2{(h_i\cdot
  h_j)\over   (x_i-x_j)^2}\big)|_{\rm  multilinear~in~h_i}$   for  the
bosonic string
and
 at $F_n=  \int   \prod_{i=1}^n  d\eta_i  \,
\exp - \sum_{i\neq j}\big({ \eta_i
\sqrt{\alpha'} (\theta_i-\theta_j)(h_i\cdot k_j)- \eta_i\eta_j(h_i\cdot
  h_j)\over (x_i-x_j+\theta_i\theta_j)}\big)$ for the superstring case
where $\eta_i$ are anticommuting variables.\vspace{-0.4cm}

\section{The four-point amplitude}

In order to understand the relations that monodromy imposes we
will begin with a discussion of the relations that arise at
four points~\cite{Plahte:1970wy,Dotsenko:1984nm}.
In that case, we can expand the amplitude
$\cA_4 \sim g_{\rm YM}^2\, \tr(T^1T^2T^3T^4) \cA{(1,2,3,4)}$ plus permutations.

For simplicity, we phrase the discussion in terms of tachyon
amplitudes. With the choice  $x_1=0$, $x_3=1$ and
$x_4=+\infty$, all three different color-ordered amplitudes
$\cA{(i,j,k,l)}$ are given by the same integrand
$|x_2|^{2\alpha'\,k_1\cdot k_2}|1-x_2|^{2\alpha'\,k_2\cdot
k_3}$ but with $x_2$ integrated over different
domains:\vspace{-0.2cm}
\begin{eqnarray}
\label{e:A1234}\!\!\!\!\!\cA{(1,2,3,4)}&=&\!\int_0^1 \!dx\,\,\,\ \, \ x^{2\alpha'\,k_1\cdot   k_2}
(1-x)^{2\alpha'\,k_2\cdot k_3}\,,\\
\!\!\!\!\!\cA{(1,3,2,4)}\!&=&\!\int_1^\infty \!\!\!dx\,\,\,\ \ \, x^{2\alpha'\, k_1\cdot k_2}
(x-1)^{2\alpha'\,k_2\cdot k_3}\,,\\
\!\!\!\!\!\cA{(2,1,3,4)}\!&=&\!\int_{-\infty}^0\!\!\!\!\! dx\,(-x)^{2\alpha'\, k_1\cdot k_2}
(1-x)^{2\alpha'\,k_2\cdot k_3}\,.
\end{eqnarray}\vskip-6pt
By a relabeling of indices, we can derive all the four-point
relations shown below from just the first of these integrals.
However, the present form is more useful for exploiting the monodromy
relations~\cite{Plahte:1970wy,Dotsenko:1984nm} satisfied by the $n$-point
amplitudes.

We first consider $\cA{(1,3,2,4)}$, where we can indicate the
contour integration from 1 to $+\infty$ by\\
\vspace{-0.2cm}
\begin{figure}[h]
\centering
\includegraphics[width=8.7cm]{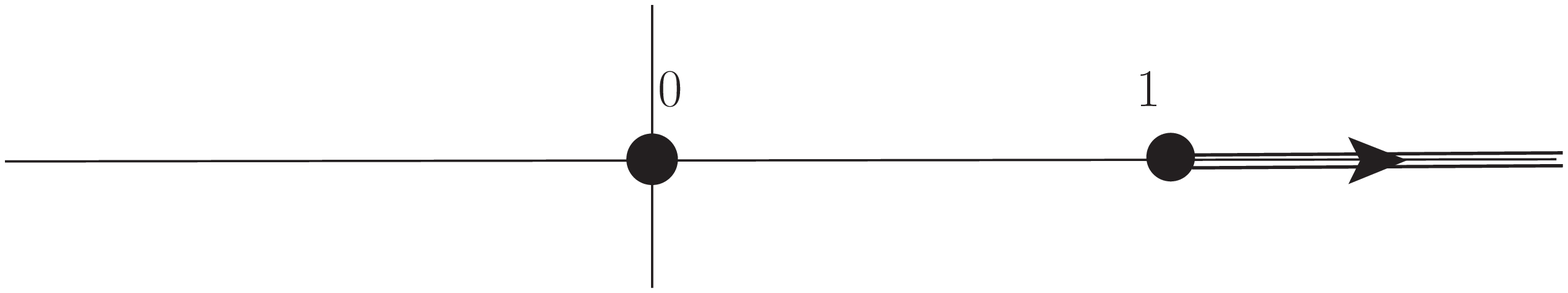}\vspace{-0.4cm}
\caption{\label{fig1}  \sl Contour
of integration for the amplitude $\cA{(1,2,3,4)}$.\vspace{-0.4cm}}
\end{figure}
Assuming that the $\alpha'\,k_i\cdot  k_j$ are complex  with
negative real parts,  we can deform the integration region so that
instead of integrating between from 1 to $+\infty$ on the real
line we integrate either on a contour slightly above or below
the real axis. By deforming each of the contours, one can
convert the expression into an integration  from $-\infty$  to
1. When rotating the contours one needs to include the
appropriate phases each time $x$ passes through $y=0$ or
$y=1$,\vspace{-0.15cm}
\begin{equation}
\label{eq:phaseP}
(x-y)^{\alpha}= (y-x)^{\alpha}\times
\begin{cases}
e^{+ i\pi \,\alpha}&\textrm{for~clockwise~rotation}\,, \cr
e^{- i\pi \,\alpha}&\textrm{for~counterclockwise~rotation}\,.
\end{cases}\vspace{-0.15cm}\nonumber
\end{equation}
The deformation of the integration region can thus be done by rotating
in the upper half plane\\
\vspace{-0.2cm}
\begin{figure}[h]
 \centering
\includegraphics[width=8.7cm]{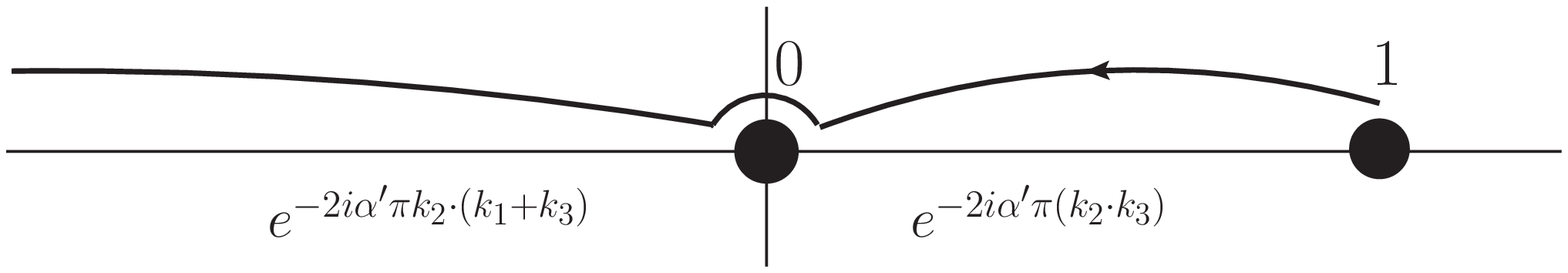}\vspace{-0.2cm}
 \caption{\sl Flipped contour for the
 amplitude $\cA{(1,2,3,4)}$.\vspace{-0.7cm}
 \label{fig2}}
\end{figure}
Because the original amplitude is real, the real part of
this contour integral expresses the original amplitude
\begin{equation}
\label{e:Re4}
\begin{split}
\cA{(1,3,2,4)}&=-\Re{\rm  e}\,\Big(  e^{-2i\alpha'\pi\,k_2\cdot  k_3}\,
\cA{(1,2,3,4)}\cr
&\hspace{0.27cm} +e^{-2i\alpha'\pi\,k_2\cdot(k_1+k_3)}\,\cA{(2,1,3,4)}\Big)\,,
\end{split}\end{equation}
where  the minus  sign arises  from  the reversed  orientation of  the
contour. The imaginary part vanishes:
\begin{equation}
\label{e:Im4}
\begin{split}
0&=\Im{\rm      m}\,\Big(      e^{-2i\alpha'\pi\,k_2\cdot      k_3}\,
\cA{(1,2,3,4)}\cr
&\hspace{0.14cm} +e^{-2i\alpha'\pi\,k_2\cdot(k_1+k_3)}\,\cA{(2,1,3,4)}\Big)\ .
\end{split}\end{equation}
This system of equations implies that all amplitudes can be related to
$\cA{(1,2,3,4)}$:
\begin{equation}
\begin{split}
\cA{(1,3,2,4)}&= {\sin(2\alpha'\pi\,k_1\cdot k_2)\over \sin(2\alpha'\pi\,
k_2\cdot k_4)}\,\cA{(1,2,3,4)}\,,\cr
\cA{(2,1,3,4)}&={\sin(2\alpha'\pi\,k_2\cdot k_3)\over \sin(2\alpha'\pi\,
k_2\cdot k_4)}\,\cA{(1,2,3,4)}\,,
\end{split}\end{equation}
where we have used momentum conservation and the on-shell
condition, $\alpha'k^2 = +1$. For other external states of
higher spin with the inclusion  of the appropriate $F_n$
factor, the integrals change in order to restore the identities
(including sign factors for the fermionic statistics of
half-integer spins). These relations are valid for  all
four-point amplitudes in bosonic and supersymmetric string
theory, as can immediately be checked using the explicit
expressions for such string amplitudes.

Taking the limit $\alpha'\rightarrow  0$, we get  the following
relations between field theory amplitudes:
\begin{equation}
\label{e:relationFour}
\begin{split}
A{(1,3,2,4)}&={k_1\cdot k_2\over k_2\cdot k_4} \,A{(1,2,3,4)}\,,\cr
A{(2,1,3,4)}&
={k_2\cdot k_3\over k_2\cdot k_4} \,A{(1,2,3,4)}\,.\vspace{-0.2cm}\
\end{split}\hspace{1.2cm}\end{equation}
These identities agree with those of ref.~\cite{Bern:2008qj}.\vspace{-0.3cm}

\section{The N-point amplitude}

We now turn to the general $n$-point case. By generalizing
the  four-point  case  we will  prove
that any  color-ordered $n$-point
amplitude can be expressed in terms of a minimal basis of $(n-3)!$ amplitudes ${\cB}$.
In the field  theory limit these  relations  reduce to  the  new amplitude
relations conjectured  in ref.~\cite{Bern:2008qj}.

First, we show to how to reduce the number of
independent amplitudes from $(n-1)!$ to $(n-2)!$ In this way we derive
a string theory generalization of the so-called Kleiss-Kuijf
relations in field theory~\cite{Kleiss:1988ne,DelDuca:1999rs}.
Indeed, in the limit $\alpha' \to 0$, our relations reduce to those,
providing an immediate and alternative proof of them.

Our starting point will be the most general amplitude, given in term
of an integral with three
fixed points, one at $0$:  $x_1=0$, one at $1$: $x_{\alpha_k}=1$, and
one at $+\infty$: $x_{n}=+\infty$. There can then be $r$
points $\{\beta_1,\ldots,\beta_r\}$ in the interval $]-\infty,0[$,
$k-1$ points $\{\alpha_1,\ldots,\alpha_{k-1}\}$ in the
interval $]0,1[$ and $s-k$ points $\{\alpha_{k+1},\ldots,\alpha_{s}\}$
in the interval $]1,+\infty[$. Both $r$ and $k$ are arbitrary, and of course
$s=n-r-2$. (We use the notation $]a,b]=\{x | a<x\leq b\}$.) We first
focus   on  the   integrations  of   the  $\{\beta_1,\ldots,\beta_r\}$
variables in the amplitude $\cA(\beta_1,\ldots,\beta_r,1,\alpha_1,\dots,\alpha_s,n)$,
illustrated in the figure below:\\\vspace{0.1cm}
\vspace{-0.3cm}
\begin{figure}[h]
 \centering
\includegraphics[width=8.7cm]{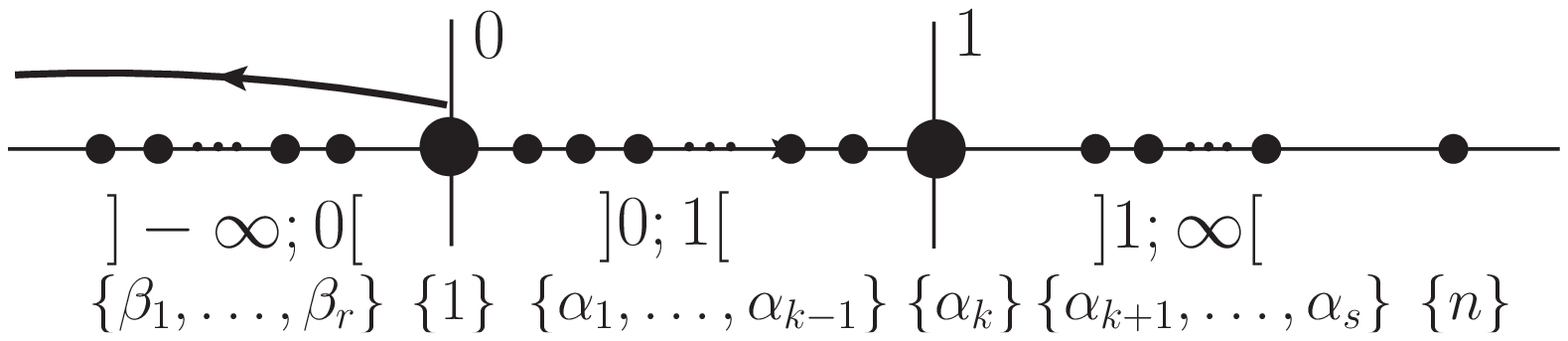}\vspace{-0.2cm}
 \caption{\sl Contour for the amplitude
 $\cA(\beta_1,\ldots,\beta_r,1,\alpha_1,\dots,\alpha_s,n)$.\vspace{-0.3cm}
    \label{fig3}}
\end{figure}
By analytic continuation of the
integration region $]-\infty,0[$
we now flip the $\beta_i$-integrations into
the region $]0,+\infty[$ in one go:\vspace{-0.05cm}\\
\vspace{-0.2cm}
\begin{figure}[h]
     \centering
     \includegraphics[width=8.7cm]{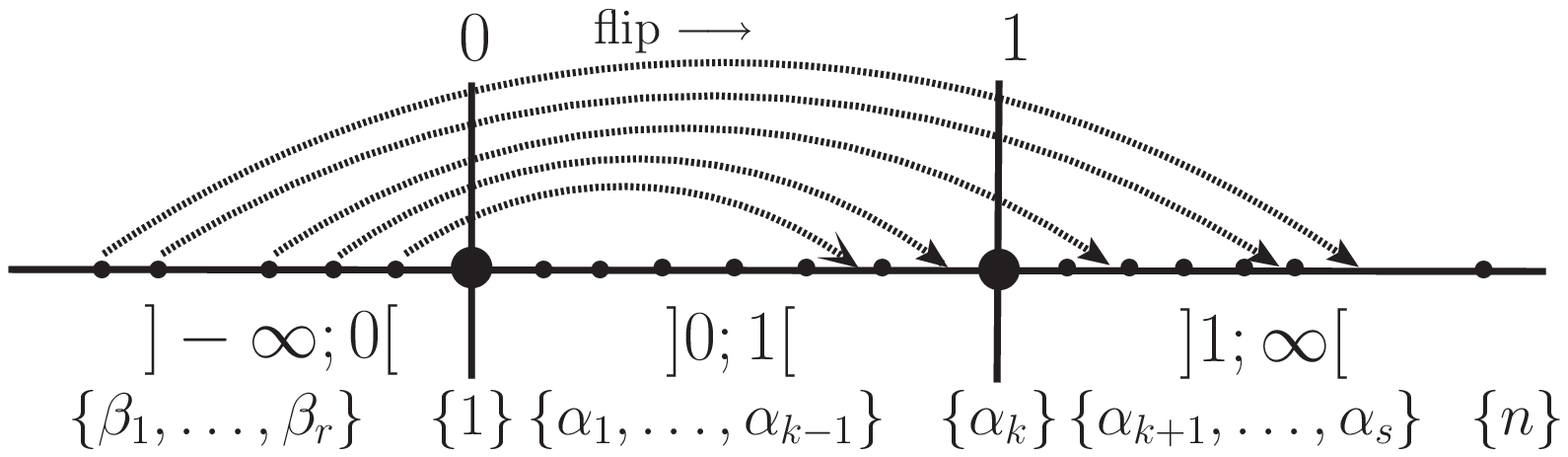}\vspace{-0.3cm}
     \caption{\sl Flipped~contour~of~fig.~\ref{fig3}.\vspace{-0.3cm}
     \label{fig4}}
   \end{figure}
We thus have an identity that relates the original integral
with integrations in the domain $]-\infty,0[$ with a sum of
integrations in the complementary region
$]0,+\infty[$.\vspace{0.35cm}
\vspace{-0.3cm}
\begin{figure}[h]
     \centering
     \includegraphics[width=8.7cm]{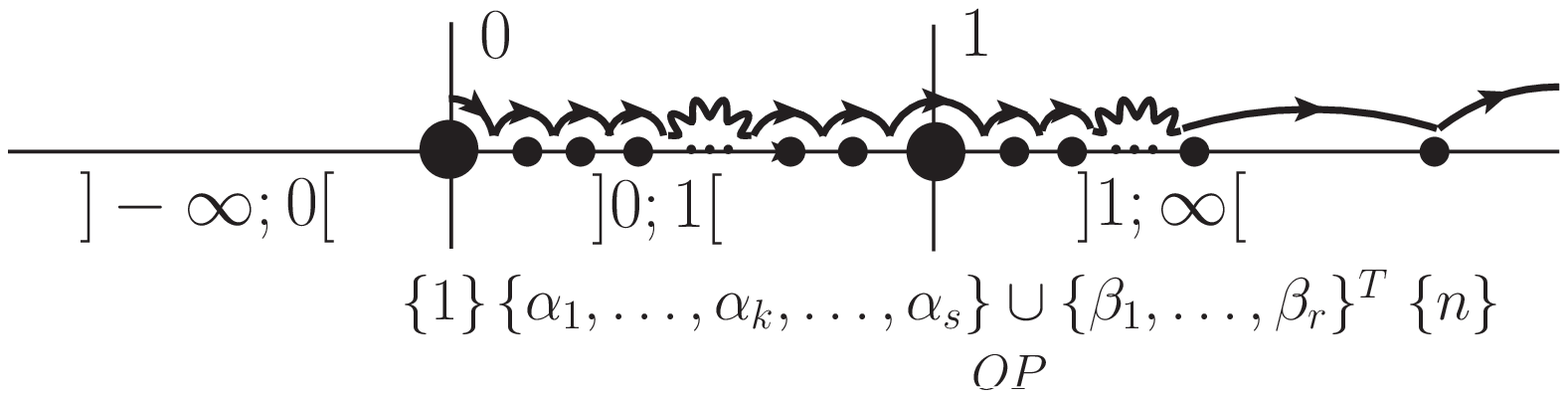}\vspace{-0.5cm}
     \caption{\sl
     Integrals~associated~with~the~contours~of~fig.~\ref{fig4}.\vspace{-0.3cm
     }
     \label{fig5}}
   \end{figure}

Taking the real parts of this $n$-point equation we arrive at
the following amplitude relation:\vspace{-0.3cm}\
\begin{eqnarray}\label{stringKK}
{\cal A}_n(\beta_1,\ldots,\beta_r,1,\alpha_1,\ldots,\alpha_s,n)&
=&(-1)^r\times\\&&\nonumber\hspace{-5.5cm}\Re{\rm e}\Big[\!\!\!\!\!\!
\prod_{1\leq i <j \leq r}\!\!\!\!\!\!\!e^{2i\pi\alpha'(k_{\beta_i}\cdot
k_{\beta_j})}\!\!\!\!\!\!\!\!\!\!\!\!\!\!
\sum_{\sigma\subset{\rm OP}
\{\alpha\}\cup\{\beta^T\}}\!\prod_{i=0}^s
\prod_{j=1}^r\! e^{(\alpha_i,\beta_j)}{\cal A}_n(1,\sigma,n)\Big],\vspace{-0.2cm}\
\end{eqnarray}\vskip-6pt\noindent
with $e^{({\alpha,\beta})} \equiv e^{2i\pi\alpha'(k_\alpha\cdot
k_\beta)}$ if  $x_{\beta} >  x_{\alpha}$  and 1  otherwise,
$\alpha_0$ denotes the leg 1 at point 0. The $(-1)^r$  arises
because the flip is reversing the $r$ integrations over the
$\beta_i$-variables. In (\ref{stringKK}) the sum runs over the
ordered set of permutations that preserves the order within
each set. These new relations between string theory amplitudes
are generalizations of the field theory Kleiss-Kuijf
relations,\vspace{-0.2cm}
\begin{equation}\begin{split}
\hspace{0cm}A_n(\beta_1,\ldots,\beta_r,1,
\alpha_1,\ldots,\alpha_s,n)&=
(-1)^r\!\!\!\!\!\!\!\!\!\!\!\!\!\!\sum_{\sigma\subset{\rm OP}
\{\alpha\}\cup\{\beta^T\}}\!\!\!\!\!\!\!\!\!\!\!
A_n(1,\sigma,n)\,,\vspace{-0.2cm}
\end{split}\end{equation}\vskip-10pt\noindent
to which they
reduce when $\alpha' \to 0$ since all phases become unity
in that limit.
The string theory relations (\ref{stringKK}) reduce the set of
independent amplitudes from $(n-1)!$ to $(n-2)!$ in detail
by eliminating all amplitudes with legs in the integration
interval $]-\infty,0[$ in favor of those with legs in the
interval $]0,+\infty[$, with the two extreme ends fixed.
However, we have
not yet used all the information contained in these $n$-point
monodromy relations.

Because the amplitudes ${\cal
A}_n(\beta_1,\ldots,\beta_r,1,\alpha_1,\ldots,\alpha_s,n)$ are
real, the imaginary parts of the $n$-point relations give
\begin{equation}\begin{split}\label{stringBern}
0=\hspace{-0cm}\Im{\rm m}\Big[\!\!\!\!\!\!
\prod_{1\leq i <j \leq r}\!\!\!\!\!\!
\!e^{2i\pi\alpha'(k_{\beta_i}\cdot k_{\beta_j})}\!\!\!\!\!\!\!\!\!\!\!\!\!\!
\sum_{\sigma\subset{\rm OP}
\{\alpha\}\cup\{\beta^T\}}\!\prod_{i=0}^s\prod_{j=1}^r
\!e^{(\alpha_i,\beta_j)}\cA_n(1,\sigma,n)\Big].\vspace{-0.3cm}\
\end{split}\end{equation}\vskip-10pt\noindent
By systematically using these relations, we can connect all
amplitudes which have points in the region $]1,+\infty[$ with
amplitudes which have points only  in the  region $[0,1]$ (and
one leg fixed at infinity).

Our proof is as follows.\! First we directly eliminate all
amplitudes with points between $]\!\!-\!\!\infty,\!0[$ in favor
of am\-pli\-tudes with legs in the interval $]0,+\infty[$,
using (\ref{stringKK}). Next using (\ref{stringBern}) we can
rewrite amplitudes of the kind ${\cal
A}_n(1,\alpha_1,\ldots,\alpha_k,
\gamma_1,\ldots,\gamma_{n-2-k},n)$ in terms of amplitudes with
at least one $\gamma_i$ among the set
$\{\alpha_1,\ldots,\alpha_k\}$ and now with at most $n-3-k$
elements between $]1,+\infty[$. For each set\!\! $\{\gamma\}$
we can find an identity in (\ref{stringBern}) so that
proceeding iteratively downward on the number of elements in
$\{\gamma\}$ starting with $n-2-k$ elements, we can thus
express all amplitudes having points in the interval
$]1,+\infty[$ in terms of $(n-3)!$ amplitudes restricted to the
interval $[0,1]$ (and one leg at infinity).

Explicitly, the five-point case gives
\begin{widetext}\vspace{-0.8cm}
\begin{eqnarray}\vspace{-0.4cm}
\nonumber \cSS{k_2}{k_5}\, \cA(2,1,3,4,5)&=& \cSS{k_2}{k_3+k_4}
\,\cA(1,2,3,4,5) + \cSS{k_2}{k_4}\, \cA(1,3,2,4,5)\,,\\
 \nonumber \cSS{k_3}{k_5}\, \cA(1,2,4,3,5)&=&
\cSS{k_3}{k_1+k_2}\,\cA(1,2,3,4,5)+\cSS{k_1}{k_3}\,  \cA(1,3,2,4,5)\,,\nonumber\\
\label{e:Sol5} \cSS{k_2}{k_5}\cSS{k_1}{k_4}\,\cA(2,3,1,4,5)&=&-\cSS{k_1}{k_2}\cSS{k_3}{k_4}
 \,\cA(1,2,3,4,5)
- \cSS{k_2}{k_4}\cSS{k_1}{k_3+k_4}\, \cA(1,3,2,4,5)\,,\\
\nonumber \cSS{k_3}{k_5}\cSS{k_1}{k_4}\, \cA(1,4,2,3,5)&=&
-\cSS{k_1}{k_2}\cSS{k_3}{k_4}\,\cA(1,2,3,4,5)
-\cSS{k_1}{k_3}\cSS{k_4}{k_1+k_2}\,\cA(1,3,2,4,5)\,,\\
 \cSS{k_1}{k_4}\cSS{k_2}{k_5}\cSS{k_3}{k_5}\,
\cA(2,1,4,3,5)&=&\Big(\cSS{k_2}{k_3+k_4}\cSS{k_3}{k_1+k_2}\cSS{k_1}{k_4}-
\cSS{k_2}{k_3}\cSS{k_1}{k_2}\cSS{k_3}{k_4}\Big)\, \cA(1,2,3,4,5) \nonumber\\
\nonumber &+&\cSS{k_1}{k_3}\cSS{k_2}{k_4}\cSS{k_5}{k_2+k_3} \, \cA(1,3,2,4,5)\,, \ \ \ \,
\end{eqnarray}
\end{widetext}\vskip-15pt
where we have introduced the notation $\cSS{p}{q}\equiv
\sin(2\alpha'\pi\,p\cdot q)$. Analogous equations are obtained
by the exchange of labels $2\leftrightarrow 3$. It is immediate
to verify these relations from the explicit form of  tree
amplitudes in string theory amplitudes in string theory given
in~\cite{Medina:2002nk,Stieberger:2006te,Boels:2008fc}. In the
field theory limit they reduce to the relations discussed in
ref.~\cite{Bern:2008qj}.\vspace{-0.3cm}

\section{Gravity amplitudes}\vspace{-0.2cm}
We finally turn towards implications of these results for
gravity amplitudes. The  $n$-point  closed  string   amplitudes
can  be  represented  as  a left/right product of color-ordered
open string amplitudes through the Kawai-Lewellen-Tye
relations~\cite{KLT}. Using the result of the previous section,
we can expand each open string amplitude of this sum in the basis of open string amplitudes
$(\cB^I,\tilde\cB^J)$:\vspace{-0.15cm}
\begin{equation}
\label{e:KLT}
\cM_n = \alpha'\,\left(\kappa\over \alpha'\right)^{n-2}\,\sum_{1\leq I,J\leq (n-3)!} \cG_{IJ}(\{k_i\})\,
\cB^I \tilde \cB^J\,.
\end{equation}\vskip-5pt
The holomorphic factorization of the  amplitude into  left and
right open string amplitudes introduces $n-3$ extra phase
factors~\cite{KLT} of the type discussed above and the entries of the matrix $\cG$ are
rational functions of degree $n-3$ in  the quantities
$\sin(2\pi\alpha'\,p\cdot q)$. Since the matrix is symmetric
this provides  a left/right  symmetric expression  for the  the
gravity amplitudes in terms  of the color ordered gauge theory
amplitudes.

As a direct application of our procedure, we can rewrite the
Kawai-Lewellen-Tye relations at four-point level
as\vspace{-0.25cm}
\begin{equation}
\cM_4 ={\kappa^2\over \alpha'}\,{\cSS{k_1}{k_2}\cSS{k_1}{k_4}\over\cSS{k_1}{k_3}}\,|\cA_4(1,2,3,4)|^2\,.
\label{KLTFour}
\end{equation}\vskip-7pt\noindent
Similarly, the five point closed string amplitude takes the
symmetric form\vspace{-0.35cm}
\begin{eqnarray}
\nonumber\cM_5         &                 =&
{\kappa^3\over {\alpha'}^2}\,\Big[\cG_{11}\,|\cA_5(1,2,3,4,5)|^2+\cG_{22} |\cA_5(1,3,2,4,5)|^2 \\
 &&\ \ \ \ \ \ +\cG_{12}
  \big(\cA_5(1,2,3,4,5) \tilde \cA_5(1,3,2,4,5)\\
&&\nonumber\ \ \ \ \ \hspace{0.78cm} +\cA_5(1,3,2,4,5)\tilde \cA_5(1,2,3,4,5)\big)\Big],
\label{KLTFive} \end{eqnarray}\vskip-9pt\noindent
where\vspace{-0.4cm}
\begin{widetext}\vspace{-0.8cm}
\begin{eqnarray}\vspace{-0.3cm}
\label{e:GG}
\cSS{k_2}{k_5}\cSS{k_3}{k_5}\cSS{k_1}{k_4}\, \cG_{11}&=&\cSS{k_1}{k_2}
\cSS{k_3}{k_4}\,\Big(\cSS{k_2}{k_3+k_4}\cSS{k_3}{k_1+k_2}
\cSS{k_1}{k_4}-\cSS{k_2}{k_3}\cSS{k_1}{k_2}\cSS{k_3}{k_4}\Big)\,,\\
\cSS{k_3}{k_5}\cSS{k_2}{k_5}\cSS{k_1}{k_4}\, \cG_{22}&=
&\cSS{k_1}{k_3}\cSS{k_2}{k_4}\Big(\cSS{k_3}{k_2+k_4}\cSS{k_2}
{k_1+k_3}\cSS{k_1}{k_4}-\cSS{k_2}{k_3}\cSS{k_1}{k_3}\cSS{k_2}{k_4}\Big)\,,\\
\cSS{k_2}{k_5}\cSS{k_3}{k_5}\cSS{k_1}{k_4}\,
\cG_{12}&=&\cSS{k_1}{k_2}\cSS{k_1}{k_3} \cSS{k_2}{k_4}
\cSS{k_3}{k_4}\,\cSS{k_5}{k_2+k_3}\,.
\end{eqnarray}
\end{widetext}\vspace{-0.2cm}
In the limit $\alpha'\to 0$ the $\cSS{p}{q}$ are replaced by
the scalar products $2\pi\, \alpha'\,(p\cdot  q)$. They lead to
an expression for  the field theory gravity amplitude that
reproduces the results  of~\cite{Bern:2008qj}. It is now clear
how this symmetric form can be proven for any number of
external states.\vspace{-0.5cm}

\section{Conclusion}\vspace{-0.25cm}

To  conclude, we  have derived  a new  series of  amplitude
identities based  on  monodromy  for  integrations in  string
theory,  providing relations between different color-ordered
amplitudes in either bosonic and  supersymmetric string theory.
As a first step, we have derived the string theory generalization
of Kleiss-Kuijf relations, thus providing a new and very simple proof
of these relations also in the field theory limit.
Our main result is the proof that there is a minimal basis of
only $(n-3)!$
amplitudes in which all other amplitudes can be expanded.  This
follows from fixing  three of the $n$
external legs at 0, 1 and $+\infty$ using the
$SL(2,\mathbb{R})$ invariance of the amplitudes, and forcing the
remaining $n-3$ coordinates to lie in the interval  $[0,1]$.
Because  the  monodromy relations  hold  for  all polarization
configurations and any  smaller number of dimensions by a
trivial dimensional  reduction, it follows immediately  that
they hold for any  choice of external legs corresponding to
the full ${\cal N} = 1$,  $D=10$  supermultiplet and dimensional
reductions thereof. The  field theory limit of  these relations
generalize and prove for any number  of external legs the new
amplitude relations recently  conjectured  by  Bern  et
al.~\cite{Bern:2008qj}  in  gauge theory.   The    string
theory   monodromy    identities   for   the Kawai-Lewellen-Tye
relationship   between  closed  and   open  string amplitudes
give  highly symmetric forms for
tree-level amplitudes between any external states in the
${\cal N}=8$, $D=4$  supermultiplet. This and other related
issues will be discussed in detail elsewhere.

{\sc Acknowledgments:}~ We thank R. Boels for comments on the manuscript
and for pointing out ref.~\cite{Plahte:1970wy}. We also thank P. Di Vecchia
for discussions and an anonymous referee for constructive comments. (NEJBB) is Knud
H\o jgaard Assistant Professor at the Niels Bohr International Academy.
The research of (PV) was supported in part by the ANR grant BLAN06-3-137168.

\end{document}